\documentclass[a4paper,11pt]{article}
\pdfoutput=1 
\usepackage{jinstpub} 
\usepackage[utf8]{inputenc}	
\usepackage[T2A]{fontenc} 
\usepackage{pdfpages}

\begin{document}
\title{Magnetic shielding for large photoelectron multipliers
for the OSIRIS facility of the JUNO detector}
\author[a]{O.\ Smirnov, \note{corresponding author,}}
\author[a]{D.\ Korablev,}
\author[a]{A.\ Sotnikov,}
\author[b]{A.\ Stahl,}
\author[b]{J.\ Steinmann,}
\author[c]{V.\ Khudyakov,}
\author[d]{I.\ Avetissov,}
\author[d]{M.\ Zykova}

\affiliation[a]{Joint  Institute  for  Nuclear  Research,  141980,  Dubna,  Russian Federation}
\affiliation[b]{III. Physikalisches Institut B, RWTH Aachen University, Aachen, Germany}
\affiliation[c]{Hydromania Ltd., 220035, Minsk, Republic of Belarus}
\affiliation[d]{Mendeleev University of Chemical Technology of Russia, Moscow, 125047, Miusskaya sq. 9, Russian Federation}

\emailAdd{osmirnov@jinr.ru}

\maketitle
\begin{abstract}

  We present technical design and characteristics of the magnetic shield developed for 20" PMTs of the low-background OSIRIS facility. A ribbon of amorphous alloy with extreme magnetic permeability was used in its design, providing excellent efficiency in screening the Earth's magnetic field with a relatively small amount of material. The mass of materials is crucial to construction of low-background facilities because of radioactive backgrounds. Using amorphous materials is cost-efficient compared to other methods for screening the Earth's magnetic field.
\end{abstract}

\section{Introduction and brief overview of methods for screening weak magnetic fields in scintillation detectors}

Normal operation of photolelectron multipliers (PMTs) is disturbed in the weak Earth's magnetic field of about 50 $\mu$T. Special measures are usually taken in order to avoid degradation of detector characteristics in large volume scintillator detectors in the weak magnetic field. The most straightforward method is compensation of the  Earth's magnetic field (EMF) at PMT locations by rather large Helmholtz coils placed outside the detector. This method was used in the Super-Kamiokande~\cite{SuperK} and KamLAND detectors~\cite{KamLAND}. The EMF compensation will be used in the JUNO detector~\cite{JUNO} in the same way. In the SuperKamiokande location, the Earth's magnetic field with a magnitude of 45~$\mu$T is oriented at 45 degrees to the horizon. Geomagnetic field compensation is provided by 26 Helmholtz coils. The average residual EMF in the detector does not exceed 5~$\mu$T with current in coils ON. As a matter of fact, only the local field at the position of a PMT is important for its proper operation, so the compensation field should be tuned correspondingly. The magnetic field at various PMT locations was
measured before the tank was filled with water.  In accordance with the technical report of KamLAND, the EMF at the underground laboratory is reduced to below 5~$\mu$T from the original 35~$\mu$T, which ensures normal operation of 20" PMTs. 

20" PMTs of the JUNO detector, both of the central
detector and of the water Cherenkov detector, lose up to 60\% of their efficiency in the non-compensated EMF. Magnetic field compensation is provided by 32 coaxial Helmholtz coils which surround the detector. The calculated residual field should not exceed 5~$\mu$T at the central detector and 10~$\mu$T at the location of PMTs of the water Cherenkov detector~\cite{JUNO2}. The PMTs are insensitive to the field of this intensity.

\begin{figure}[!htb]
\centering
\includegraphics[width=0.5\textwidth]{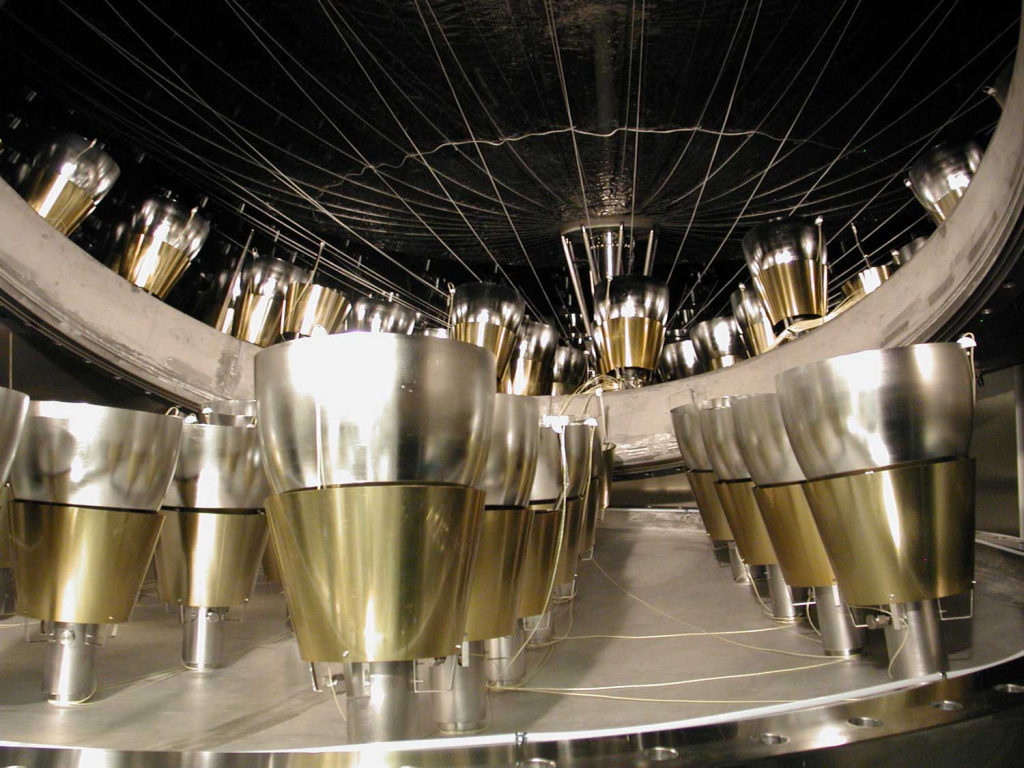}
\caption{PMTs of the Borexino detector are equipped with magnetic shields (yellow truncated cones at the PMT base in the photo, the colour is due to waterproof coating) and light concentrators. PMTs are not visible in the photo. The highest point of the PMT photocathode touches the plane of the upper edge of the magnetic shield. Courtesy of the Borexino collaboration.}
\label{BrxFig}
\end{figure}

Should EMF compensation be for some reasons unsuitable, PMT screening with high magnetic permeability is needed. The common solution is a cylinder or truncated cone made of permalloy sheets placed coaxial to PMTs. This scheme was applied to the Borexino prototype detector CTF~\cite{CTF} and Borexino detector~\cite{Brx}. The magnetic shield of the Borexino detector is shown in Fig.~\ref{BrxFig}.
    
The results of magnetic shield testing for CTF/Borexino were presented in~\cite{BrxShield}. "Open" screens are quite effective in reducing the EMF component in the direction perpendicular to the axis. Along the axis, the reduction of the field is rather moderate because of the open structure. In general, the magnetic field penetrates into the open structure at a depth comparable to the size of the hole, and the usual recommendation is to extend the magnetic shield about one diameter beyond the cathode plane~\cite{Photonis}. It is impossible to follow the recommendation in a straight way as the shadow of such a size will cut a significant part of the light field at the OSIRIS facility. Nevertheless, due to the fact that PMTs with a linear focussed dynode structure are less sensitive to the field along the axis~\cite{BrxShield}, a shorter "open" shield is acceptable for application to PMTs with a linear focussed dynode structure.

In order to suppress the EMF more uniformly, a number of experiments used cages made of permalloy wires: DarkSide-50~\cite{DS50}, Antares~\cite{Antares}, Baikal-GVD~\cite{Baikal}. The use of a non-transparent material in front of the photocathode leads to partial shadowing of light, reducing the total photoelectron detection efficiency (PDE). In practice, the loss is about~5\%. Moreover, it is impossible to achieve a significant magnetic field reduction with a cage, the field is reduced by a factor of 2.0-2.5. A common serious disadvantage of using permalloy screens (both types: either made of sheets or wire), is non-stability of shielding efficiency. Permalloy screens require annealing for achieving a high magnetic permeability. Magnetic properties of the shield can degrade because of mechanical stress. In this respect, cages are extremely sensitive to conditions of transportation and treatment - any mechanical deformation can lead to a partial loss of magnetic properties.
    
To make the description complete, let us mention a scheme of the EMF partial compensation for PMTs with a linear focussed dynode structure proposed in~\cite{Orientation}. The scheme is based on different sensitivities of PMTs to two mutually orthogonal directions (x-axis and y-axis, perpendicular to z, the PMT axis of symmetry
). Orienting PMTs by rotation around the z-axis, it is possible to minimize the projection of the EMF in a single, most sensitive direction. In such a way, the EMF influence is minimized without using screens or Helmholtz coils. Naturally, two less destructive EMF components remain uncompensated.

\begin{figure}[!htb]
\centering
\includegraphics[width=0.5\textwidth]{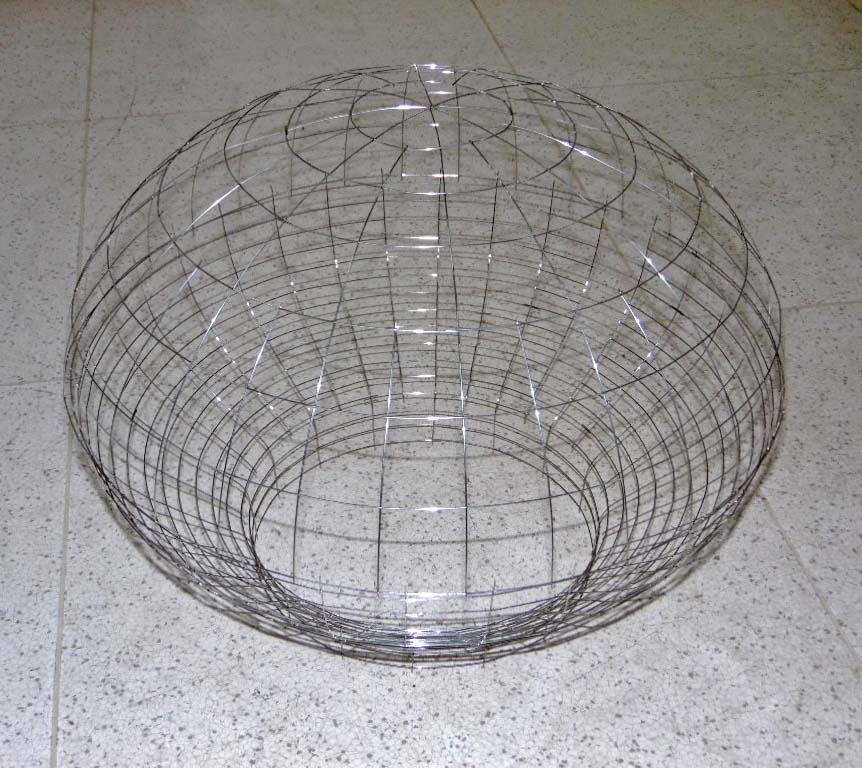}
\caption{A prototype cage for a large PMT (20"). It consists of two separate parts. The diameter of the widest part (nominal "equator") is approximately 50 cm. Permalloy wire 1 mm in diameter was used, the size of a single cell in front of the photocathode is~50 mm. The wiring below the equator is denser by a factor of 3.}
\label{GridPrototype}
\end{figure}

When being used in low-background detectors, the magnetic shield should not significantly contribute to the total radioactive background. In practice, the radioactive background of the shield should be lower than that of a PMT. From this point of view, cages are less radioactive compared to full-metal shields because of a lower amount of material used. Nevertheless, modern PMTs with a photocathode 20" in diameter will need some kilograms of material to provide reasonable screening. In Fig.~\ref{GridPrototype}, we present, as an example, one of the prototype cages developed at our Laboratory within JUNO activities  for 20" Hamamatsu PMTs (type R12860).
    
Permalloy wire 1 mm in diameter was used for the cage with a ~50 mm cell, the light loss on the mesh is about 5\%\footnote{Because of photocathode sphericity and dependence of quantum efficiency on the angle of incidence, the loss of light does not coincide with geometric shadowing.}.  The cages provided an acceptable reduction of the magnetic field by a factor of 2, but the annealing process was complicated because of the cage size - specialized large-volume ovens are quite rare, and waiting for processing reaches several months. Moreover, the cages need special coverage with protecting lacquer because PMTs will be placed in water. This also causes technical issues as one needs "cold" covering in order to preserve magnetic properties after annealing. The JUNO collaboration decided for a global compensation of the EMF within the detector volume, but the idea of magnetic screens came back to the agenda during discussions about the OSIRIS facility design developed for measuring ultra-low radioactivity of JUNO liquid scintillator before filling the main JUNO detector. It is impossible to use Helmholtz coils at the OSIRIS facility at the Jiangmen Underground Neutrino Observatory (JUNO) because of the restricted space. 

As the experience with large-size cages was unsatisfactory, we decided to use an open structure magnetic shield similar to the one used in the Borexino experiment but manufactured from modern materials: amorphous alloy and composite materials. The PMTs of the OSIRIS facility will be immersed in water, and hence all the materials and water should remain unchanged for a rather long period. At the same time, the materials used should not contribute much to the total radioactivity of the setup.

In the present article, we describe the successful application of the amorphous alloy ribbon for production of EMF shields for the PMTs of the low-background OSIRIS facility. The shields are made of glass fibre and/or carbon fibre composite material and are coated outside with black gelcoat providing necessary waterproof properties to the construction. The matted black colour of the covering excludes parasitic reflections. The shields are equipped with an auxiliary aluminium foil layer serving as a radiofrequency screen. In order to satisfy the demands for low radioactivity, all the materials were carefully selected based on the results of measurements.

An important advantage of using amorphous alloy is a significant decrease in the mass and cost of material while providing the same screening effect: magnetic permeability of non-annealed~\footnote{Annealing could further increase magnetic permeability by a factor of 2~\cite{Private}, but it would be technically complicated in our case.} amorphous alloy achieves $10^{6}$ compared to the typical magnetic permeability of annealed permalloy of $10^{4}$, the price is roughly an order higher. Replacing permalloy with the amorphous tape in such a way, we gain two orders of magnitude in mass and an order of magnitude in price for the same screening coefficient.

In this paper, we present a report based on the experience of manufacturing 64 shields using carbon fibre composite for the PMTs of the main detector of the OSIRIS facility with very strict requirements for material radioactivity and 12 shields for the PMTs of the muon veto detector with less strong requirements for radioactivity.

At the time our paper was being prepared, an article about a similar approach to magnetic shielding using amorphous alloys has been published~\cite{LHAASA} describing magnetic screens for 20" PMTs with MCP plates for the LHAASA-WCDA experiment. The screens surrounding the photocathode, have a cylindrical shape and are 41 cm in height. The protection against oxidation in water is provided by a polymer film. The main difference from our shields is the absence of any demands on screen radioactivity. Another significant difference is that we use the internal surface of the shield as a light concentrator.
    
\section{OSIRIS facility}

\begin{figure}[!htb]
\centering
\includegraphics[width=0.5\textwidth]{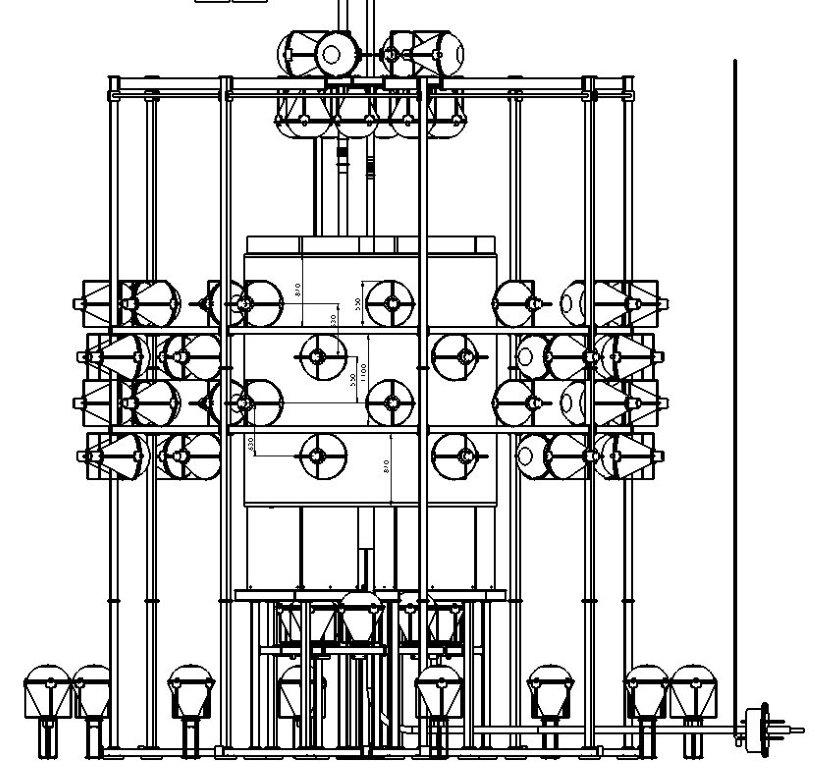}
\caption{A schematic view of the OSIRIS facility. The side projection of the central cylindrical detector is represented by a rectangular shape in the centre of the image. The active volume is viewed either by the PMTs horizontally placed around the active volume or by ones vertically installed at the bottom and at the top of the cylinder. Muon veto PMTs are installed on the "floor" (8 PMTs) or on the "ceiling" (4 PMTs) of the detector. Courtesy of the OSIRIS collaboration.}
\label{FigOSIRIS}
\end{figure}

The OSIRIS facility is schematically shown in Fig.~\ref{FigOSIRIS}. The facility uses 64 PMTs to detect light from the central liquid scintillator detector; the other 12 PMTs are used as the muon veto (water Cherenkov detector), 8 of them are installed in water at the bottom of the detector, and 4 are placed at the top of the detector.
As one can see from Fig.~\ref{FigOSIRIS}, the PMTs of the central detector are oriented either vertically, below and above the cylindrical sensitive volume, or horizontally around the central detector. 8 PMTs of the muon veto at the bottom are oriented vertically and 4 PMTs placed at the top are oriented horizontally.

The JUNO detector and its facility, OSIRIS, are under construction at the Jiangmen Underground Neutrino Observatory (JUNO) in Kaiping, China. The intensity of the geomagnetic field in the laboratory location (22.1$^{\circ}$ North) is 45~$\mu$T: 38 $\mu$T in the horizontal direction to geographic north and 24~$\mu$T in the vertical direction. The inclination of the field is 32.0$^{\circ}$ to the horizon. Because of the stronger horizontal component of the field at the laboratory compared to the vertical one, open shields effectively suppress the most sensitive component of the field (orthogonal to the PMT axis) for the PMTs vertically installed. The factor of reduction is the same for all PMTs. For the PMTs installed horizontally, the screening factor depends on orientation: efficiency of reduction of the field component parallel to the PMT axis will depend on the angle between the PMT axis and the horizontal projection of the field. 

\begin{figure}[!htb]
\centering
\includegraphics[width=0.55\textwidth]{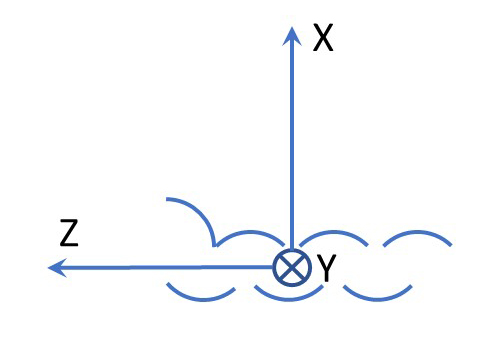}
\caption{The coordinate system for characterization of PMT sensitivity to the magnetic field is linked to the linear dynode structure in a way schematically presented in the drawing. The y-axis is directed from the observer.}
\label{Axes}
\end{figure}

The Online Scintillator Internal Radioactivity Investigation System (OSIRIS) is a 20-ton liquid-scintillator detector for monitoring radiopurity of liquid scintillator during the JUNO filling phase. The OSIRIS will operate 20" Hamamatsu PMTs (type R15343). This PMT type uses an electron multiplier with a linear focussed structure (Box and linear focussed)~\cite{OSIRIS}, also see datasheet R15343. The sensitivity of characteristics of PMTs of this type to components of the field orthogonal to the PMT axis is noticeably stronger than to those directed along the PMT axis (see.~\cite{BrxShield}). For the simplicity of notations in the following discussion, we will use the notation z for the axis along the PMT axis and x and y for two other orthogonal axes. For graphic representation of the coordinate system, see Fig.~\ref{Axes} .

Developing a magnetic shield design, we decided to use the inner part of the shield protruding from the PMT equator as a light concentrator. The height of the cylindrical part of the shield was chosen on the basis of Monte Carlo studies. The main parameter under study was efficiency of collection of light from the central detector: a cone being too high would shadow a fraction of incident light arriving at large angles at the PMT photocathode.

\section{Description of the magnetic shield for PMTs of the OSIRIS facility}

\begin{figure}[!htb]
\centering
\includegraphics[width=0.5\textwidth]{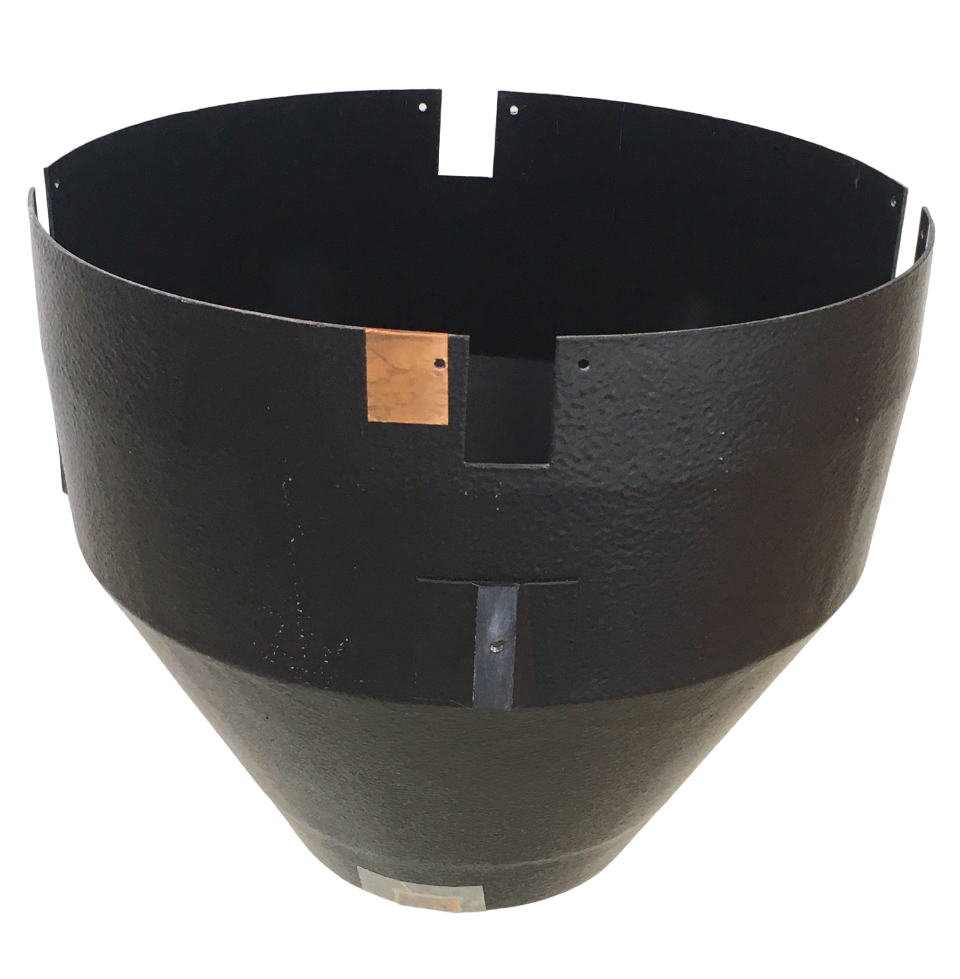}\includegraphics[width=0.5\textwidth]{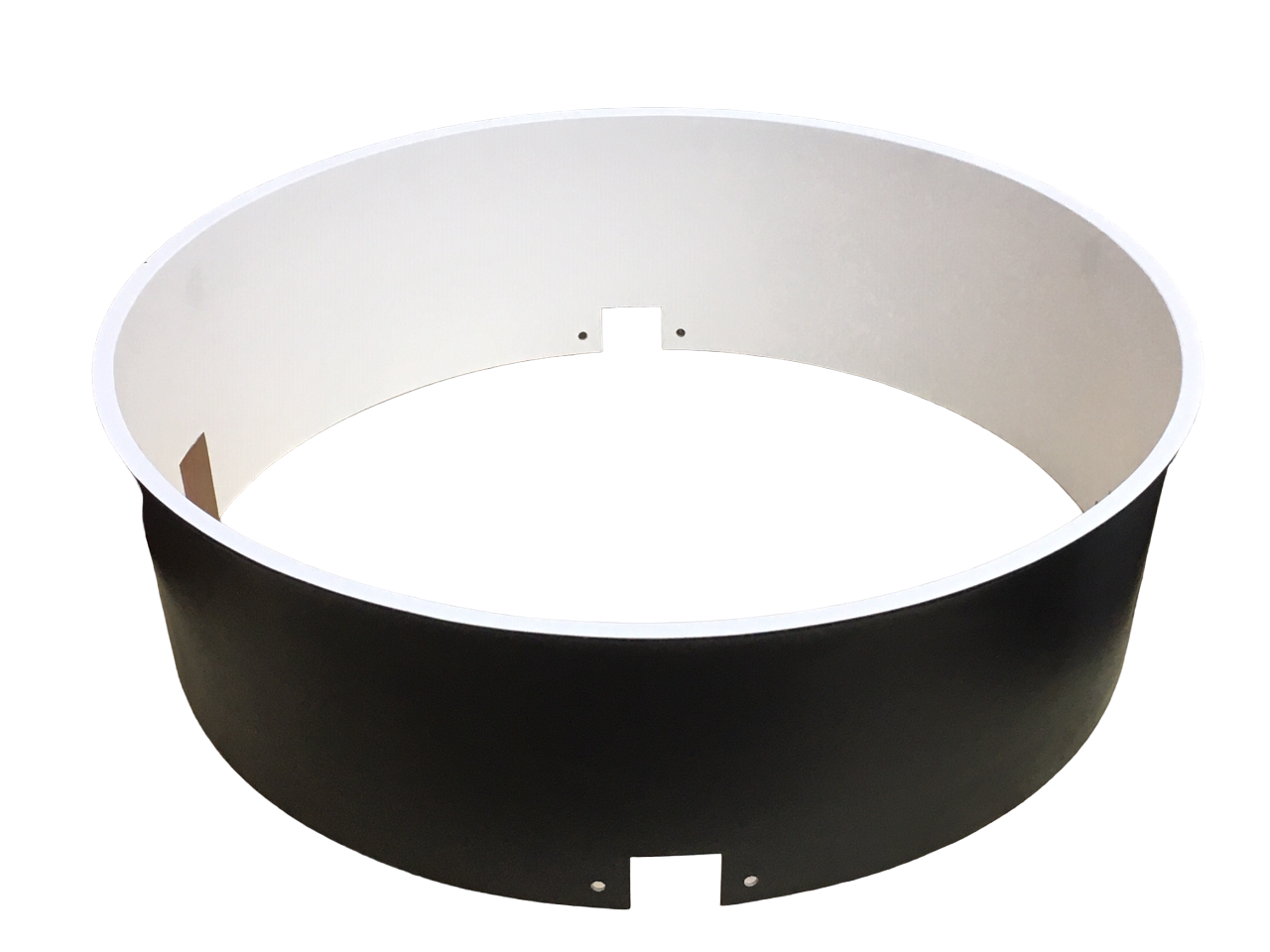}
\caption{Photos of the magnetic shield prototype: the main (left) and auxiliary (right) elements of the magnetic shield are shown separately. A copper contact of the RF screen is visible at the top of the main element. The auxiliary element is equipped with the corresponding contact, too. In the final design, the copper contact was replaced by a food grade steel in order to avoid corrosion when placed in high-purity water.}
\label{ScreenFotos}
\end{figure}

\begin{figure}[!htb]
\centering
\includegraphics[width=0.55\textwidth]{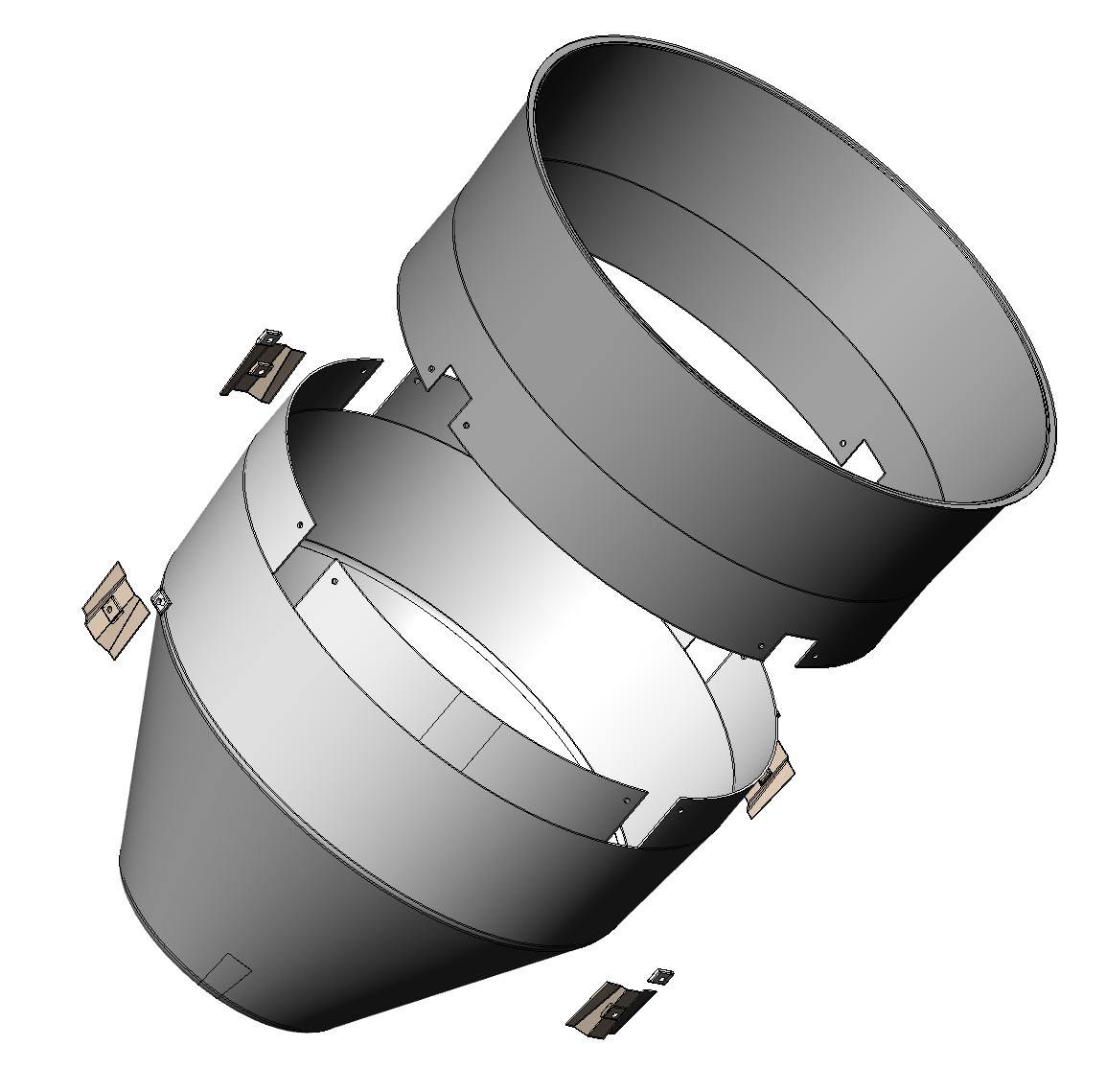}\includegraphics[width=0.45\textwidth]{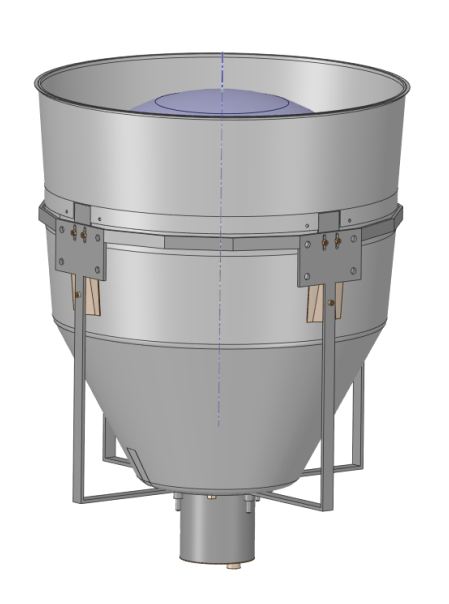}
\caption{On the left: two elements of the magnetic shield are shown in an exploded view. On the right: the magnetic shield is shown together with a PMT mounted. The main element of the magnetic shield is placed below the "equator" of the PMT bulb. The auxiliary element is installed above the "equator".}
\label{Exploded}
\end{figure}

The magnetic shield consists of two parts: the main conic part and the auxiliary cylindrical one, see Figs.~\ref{ScreenFotos}  and  \ref{Exploded}. 


The magnetic shield is formed by using the AMAG-170 ribbon
(manufactured by PC MSTATOR, Borovichi, Russia). The amorphous soft magnetic alloy used in the production of the AMAG-170 ribbon has extreme magnetic permeability of an order of $10^{6}$. The high magnetic permeability of the ribbon allows using relatively small amounts of material, and this is of great importance for the OSIRIS facility sensitive to radioactivity of construction materials. Another important feature of the amorphous soft magnetic alloy is the stability of its properties. The material does not require mandatory annealing to restore its magnetic properties after processing.

Both parts of the magnetic shield are made from composite materials (carbon fibre and glass fibre), 8 layers of the AMAG-170 ribbon $\simeq25$~$\mu$m thick and 30~mm wide (both dimensions are maximal available) are placed between the layers of composite material in a way to guarantee an average thickness of the shield of ~0.2 mm. The uniformity of tape placement does not play a big role, the total screening is defined mainly by the total mass of the material used.
One shield contains about 1.5 kg of amorphous tape.

An additional layer of aluminum foil is laid above the soft magnetic tape, it serves as a screen against electric fields (radio frequency noise). Amorphous metal does not provide effective RF screening because of rather moderate conductivity.

Two shield variants were produced: one for protection of the muon veto PMTs (12 in total) and another for shielding the PMTs of the main detector (64 in total). The main difference is the type of the composite materials used. The muon veto shields are produced using less expensive glass fibre composite materials. Glass fibre is not appropriate for the main detector in respect to radioactivity because of the high content of potassium (in this case from U and $^{40}$K), so a more expensive carbon fibre composite material on the basis of potassium-free polyether resin was used. This material provides the matted black colour. The choice of polyether resin results from its long-term resistance to chemically ultrapure water. Masses of the material used for both shield variants are presented in Tables \ref{TableWeight1} and \ref{TableWeight2}.

Glass fibre shields are covered with black gelcoat to prevent soaking of construction materials when placed in water. The black matted colour of the protection excludes parasitic reflection of the light from surfaces. The carbon fibre composite is originally black. It was covered with a transparent gelcoat (topcoat).
The auxiliary cylindrical part is covered from inside with white gelcoat. Titanium dioxide was used as a filler (10\% TiO$_{2}$ by mass). The outside surface is covered with black gelcoat and matted in the case of glass fibre shields and left untouched for carbon fibre items. In accordance with Monte Carlo simulations, we expected an about 20\% light collection improvement in the air. Measurements showed a 19\% improvement in light collection efficiency (see Section~\ref{SectionEfficiency}).

\section{Radioactivity of the magnetic shield}

The purpose of the OSIRIS facility is low-background measurements of the extremely low radioactivity levels of the liquid scintillator before filling in the main JUNO detector. The levels needed for the JUNO physics programme are at a level of $10^{-17}$-$10^{-16}$ g/g in U, Th and K and are unreachable with common methods. The flux from radioactive impurities in magnetic shield materials should not exceed the corresponding flux from PMTs. 

The analysis of chemical impurities was performed at the Mendeleev University of Chemical Technology of Russia by inductively coupled plasma mass spectrometry (ICP-MS) with preliminary transfer of solid samples to the liquid phase. Extra pure water (AquaMax-Ultra 370 Series, Young Lin Instruments Co. Ltd., Korea,  Anyang) with a resistance of 18 M$\Omega$·cm was used. Aluminum foil, steel ribbon and carbon plastics were dissolved in high-purity nitric (HNO$_{3}$) acid (7N7) purified by the Berghof BSB-939-IR surface distillation system in the SPEEDWAVE-FOUR microwave decomposition system (BERGHOF GmbH \& Co. KG, Germany, Drolshagen) equipped with DAP-100 PTFE autoclaves (BERGHOF GmbH \& Co. KG, Germany, Drolshagen). 
Polymer materials were dissolved in high-purity sulfuric (H$_{2}$SO$_{4}$) acid (8N Ultrapur, Sigma Aldrich) by using the same microwave decomposition system.
Analytical measurements were carried out with the NexION 300D inductively coupled plasma mass spectrometer (PerkinElmer Inc., USA, MA, Waltham).
 The quantitative analysis of Th and U was carried out using the “additives” method, taking into account the concentration of the main (matrix) elements in the solution analyzed. Standard solutions (PerkinElmer Inc.) were used for calibration. The optimized operating mode of the NexION 300D spectrometer applied to the impurity analysis of samples is presented in Table~\ref{TableOpMode}.
 
\begin{table}
\begin{tabular}{c|c}
\hline 
Nebulizer type & Concentric (Meinhard), PFA \tabularnewline
\hline 
Spray chamber & Scott double-pass chamber, PFA \tabularnewline
\hline 
Argon flow rate, L/min through the nebulizer & 0.98 \tabularnewline
\hline 
plasma-forming & 15 \tabularnewline
\hline 
auxiliary & 1.2 \tabularnewline
\hline 
Generator power, W & 1500 \tabularnewline
\hline 
Number of scan cycles & 15 \tabularnewline
\hline 
\end{tabular}

\caption{The operating mode of the NexION 300D instrument for conducting the impurity analysis of samples.}
\label{TableOpMode}
\end{table}

The results of measurements recalculated to the content of radioactive 
impurities are shown in Tables~\ref{TableRad1} and~\ref{TableRad2}. Note that only preselected materials which satisfy radioactivity requirements are listed in the Tables. We did not test food grade stainless steel of the electric contact because of the low mass of the material used. Even the worst case of natural content would not harm the total balance of radioactivity. One can see (Table~\ref{TableRad1}) that the glass fibre shield exceeds the reference values by an order of magnitude in U and by a factor of 4 in potassium, that is why it is not appropriate for the main detector. 

\begin{table}

\begin{tabular}{|c|c|c|c|c|}
\hline 
Material  & Manufacturer & Main  & Auxiliary  & Total\tabularnewline
          &              & element  & element  & \tabularnewline
\hline 
\hline 
Glass fibre & ECC Interglas 92140, aero & 0.8  & 0.36  & 1.16 \tabularnewline
\hline 
Vinylester resin & Derakane Momentum$^{TM}$ 470-300 & 1.018  & 0.458  & 1.476 \tabularnewline
\hline 
Black gelcoat  & Maxguard™ GN-NTRL S + carbon black & 1.4  & 0.35  & 1.75\tabularnewline
\hline 
White gelcoat & Maxguard™ GN-NTRL S + 10\% TiO$_2$& -  & 0.35 & 0.35 \tabularnewline
\hline 
Amorphous alloy & MSTATOR AMAG-170 & 1.015 & 0.55  & 1.565\tabularnewline
\hline 
Aluminum foil  & & 0.06  & 0.03  & 0.09\tabularnewline
\hline
Steel ribbon  & 12х18н10т  type & 0.03  & 0.02  & 0.05 \tabularnewline
\hline 
Total  & &4.343  & 2.128  & 6.471 \tabularnewline
\hline 
\end{tabular}

\caption{Weight of materials used in kg (glass fibre items with black gelcoat coating). We used steel Type 12х18н10т made in Russia, its western analogue is AISI 304.}
\label{TableWeight1}
\end{table}

\begin{table}
\begin{tabular}{|c|c|c|c|c|}
\hline 
Material  & Manufacturer & Main  & Auxiliary  & Total\tabularnewline
          &              & element  & element  & \tabularnewline
\hline 
\hline 
Carbon fibre  & G. Angeloni GG245T & 0.57  & 0.25  & 0.82 \tabularnewline
\hline 
Vinylester resin & Derakane Momentum$^{TM}$ 470-300 & 0.82  & 0.37 & 1.19 \tabularnewline
\hline 
Transparent topcoat  & Maxguard™ GN-NTRL S& 1.26  & 0.32  & 1.58\tabularnewline
\hline 
White gelcoat & Maxguard™ GN-NTRL S + TiO$_2$&-  & 0.35 & 0.35 \tabularnewline
\hline 
Amorphous alloy & MSTATOR AMAG-170 & 1.015 & 0.55  & 1,565\tabularnewline
\hline 
Aluminum foil  & & 0.06  & 0.03  & 0.09\tabularnewline
\hline
Steel ribbon & 12х18н10т type & 0.03  & 0.02  & 0.05 \tabularnewline
\hline 
Total  & & 3.755 & 1.890  & 5.645\tabularnewline
\hline 
\end{tabular}

\caption{Weight of materials used in kg (carbon fibre items with transparent coating).}
\label{TableWeight2}
\end{table}

\begin{table}
\begin{tabular}{|c|c|c|c|c|c|c|}
\hline 
Material & U & Th & K & m(U) & m(Th) & m($^{40}$K)\tabularnewline
 & ppb & ppb & ppm & $\mu$g  & $\mu$g  & $\mu$g \tabularnewline
\hline 
\hline 
Glass fibre & 5833  & \textless 3  & 1460 & 6770  & \textless 4  & 198\tabularnewline
\hline 
Vinylester resin  & \textless 0.1  & 0.9  & 0.78  & \textless 0.15  & 1.3  & 0.13\tabularnewline
\hline 
Black gelcoat  & \textless 3  & \textless 0.3  & 177  & \textless 5.3  & \textless 0.5  & 36\tabularnewline
\hline 
White gelcoat  & 7  & 7  & 4.33  & 2.5  & 2.5  & 0.2\tabularnewline
\hline 
AMAG-170 ribbon  & 3  & \textless 5  & 0.84  & 4.7  & \textless 7.8  & 0.154\tabularnewline
\hline 
Aluminum foil  & 170  & 26  & 0.96  & 15.3  & 2.4  & \textless 0.01\tabularnewline
\hline 
Total  & - & - & - & 6798  & \textless 18  & 234 \tabularnewline
\hline 
PMT glass  & 400  & 400  & 60 & 3600  & 3600 & 63\tabularnewline
\hline 
\end{tabular}
\caption{Abundances (three first columns) and the total content (in $\mu$g, three last columns) of natural long-lived radioactive isotopes in construction materials of the glass fibre shield. The last row presents corresponding reference values for glass with a total mass of 9 kg. The total radioactivity of glass is $\simeq$80 Bq. The total radioactivity of the glass fiber magnetic shield is  $\simeq$150 Bq, almost twice the radioactivity of the glass.}
\label{TableRad1}
\end{table}

\begin{table}
\begin{tabular}{|c|c|c|c|c|c|c|}
\hline 
Material & U & Th & K & m(U) & m(Th) & m($^{40}$K)\tabularnewline
 & ppb & ppb & ppm & $\mu$g & $\mu$g & $\mu$g\tabularnewline
\hline 
\hline 
Carbon fibre  & 3  & \textless 5  & 0.84  & 0.25  & \textless 4.9  & 1.4\tabularnewline
\hline 
Vinylester resin  & \textless 0.1  & 0.9  & 0.78  & \textless 0.12 & 0.11  & 0,1\tabularnewline
\hline 
Transparent topcoat  & \textless 3  & \textless 0.3  & \textless4.33  & \textless 4.8  & \textless 0.5  & \textless6.8\tabularnewline
\hline 
White gelcoat  & 7  & 7  & 4.33  & 2.5  & 2.5  & 0.2\tabularnewline
\hline 
AMAG-170 ribbon  & 3  & \textless 5  & 0,84  & 4.7  & \textless 7.8  & 0,154\tabularnewline
\hline 
Aluminum foil  & 170  & 26  & 0.96  & 15.3  & 2.4  & \textless 0.01\tabularnewline
\hline 
Total  & - & - & - & \textless27.7  & \textless 18  & \textless8.7 \tabularnewline
\hline 
PMT glass  & 400  & 400  & 60 & 3600  & 3600 & 63\tabularnewline
\hline 
\end{tabular}\caption{Abundances (three first columns) and the total content (in $\mu$g, three last columns) of natural long-lived radioactive isotopes in construction materials of the carbon fibre shield. The last row presents the corresponding reference values for glass with a total mass of 9 kg. We used the best value of those measured for black and white gelcoats as an upper limit for transparent topcoat. Due to the absence of fillers, the transparent topcoat should be less radioactive, but even with these values, we are still well below the levels imposed by glass radioactivity. The total radioactivity of the magnetic shield is below 3 Bq, which is negligible compared to the radioactivity of 80 Bq of the glass of the PMT.}
\label{TableRad2}
\end{table}

\section{Manufacture of the magnetic shield}

Elements of the magnetic shield were manufactured by Hydromania Ltd (Minsk, Belarus). The production line was placed in a dust-free isolated room, and all the measures were taken to prevent accumulation of any impurities on surfaces. Immediately after manufacturing, the shields were sealed in polyethylene bags to prevent contact with air and, correspondingly, contamination of surfaces with products of radon decay~\footnote{About 0.1\% of styrene used for polymerization of gelcoat (20 grams of 20\% solution of styrene in parafene were used) remains unbounded and evaporates. This process caused the accumulation of gaseous styrene in the sealed plastic bag, and a strong smell was noticed after unpacking. With the initial content of styrene in rough materials of about 4 g for a carbon fibre shield (by a factor of 6 more for one of glass fibre), about 4 mg per shield of residual styrene could emanate. When working in a large workshop, the concentration of styrene remains within safe levels (the legal airborne permissible exposure limit is 100 ppm averaged over an 8-hour workshift). Anyway, the work was done using special masks for respiratory protection. Before the final washing, the items were left in climatic boxes for removal of gaseous styrene.}. The shields will be washed with pure water before the installation in the OSIRIS facility.

The shields are produced using the technology of manual (contact) formation on mandrels. The mandrels were made of waterproof plywood and coated with a protective glass fibre layer and instrumental (matrix) gelcoat. After hardening, the gelcoat surface was polished to the necessary grade of roughness.

\begin{figure}[!htb]
\centering
\includegraphics[width=0.55\textwidth]{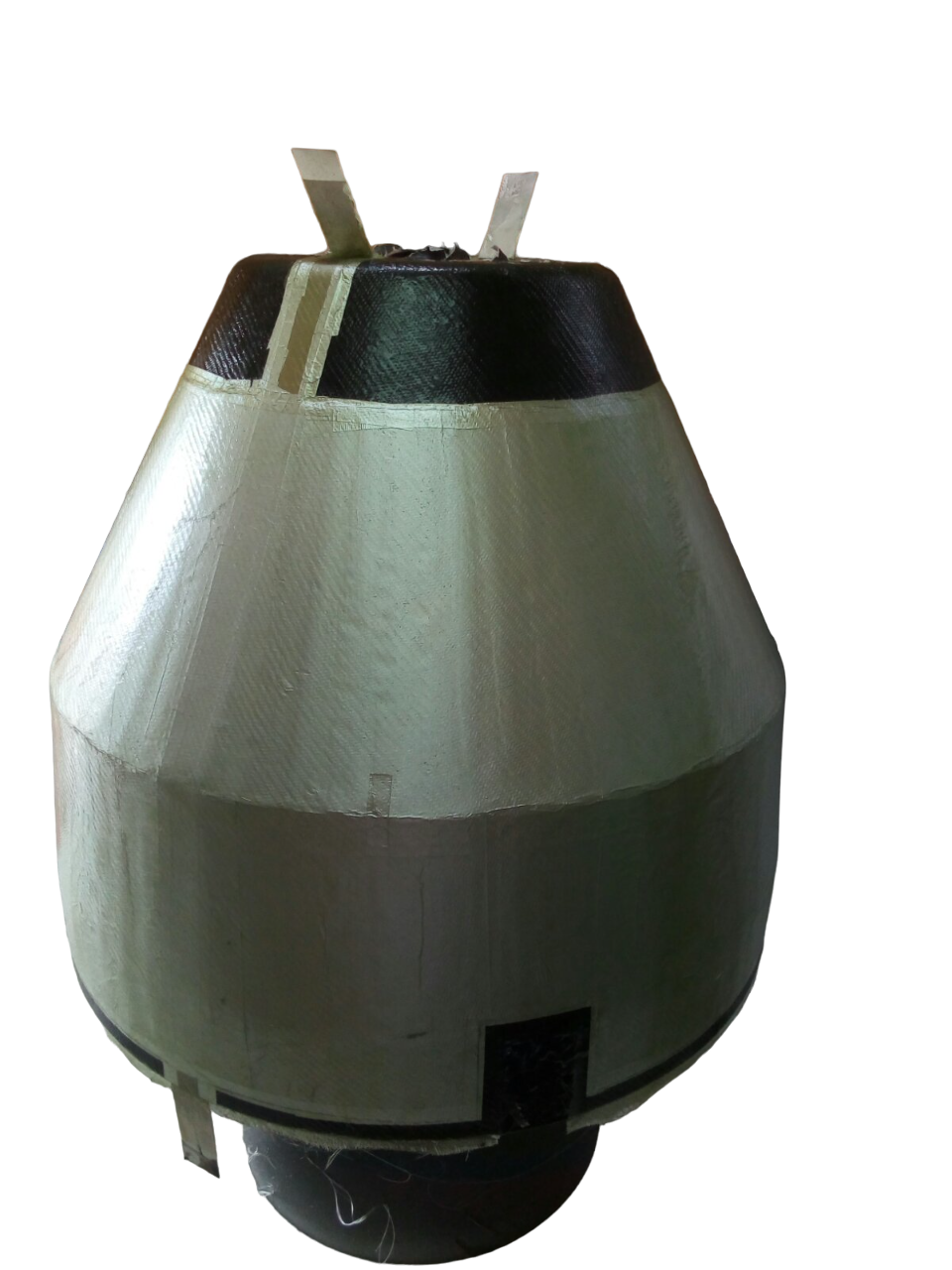}
\caption{The view of the magnetic shield before covering with outside protective layers. The location of the amorphous alloy in the shield is clearly visible, there was no need in magnetic shielding of the narrow part of the cone since this part of the PMT is insensitive to magnetic fields.}
\label{Ribbon}
\end{figure}

The sequence of technological operations for carbon fibre items was as the following:
\begin{itemize}
    \item coating of the surface with gelcoat: transparent gelcoat was used for the main element and white gelcoat for the auxiliary one;
    \item inner layers of carbon fibre with resin are placed after partial polymerisation of gelcoat (to a tack condition);    
    \item four layers of the amorphous ribbon are placed on the carbon fibre surface. The orientation of the ribbon in layers was chosen to avoid empty spaces in the shield;
    \item aluminum foil and stainless steel lamellas are placed on layers of the amorphous ribbon; the outside electric contact is formed by the protruding stainless steel lamella part;
    \item four last layers of the amorphous ribbon are placed above aluminum foil, the result is shown in Fig.\ref{Ribbon};
    \item  outer layers of carbon fibre were placed after the installation of magnetic shields and RF screens;   
    \item  finally, the outer surfaces were coated with a transparent gelcoat (topcoat). 
\end{itemize}

\section{Efficiency of the magnetic shield}
\label{SectionEfficiency}

\begin{figure}[!htb]
\centering
\includegraphics[width=0.49\textwidth]{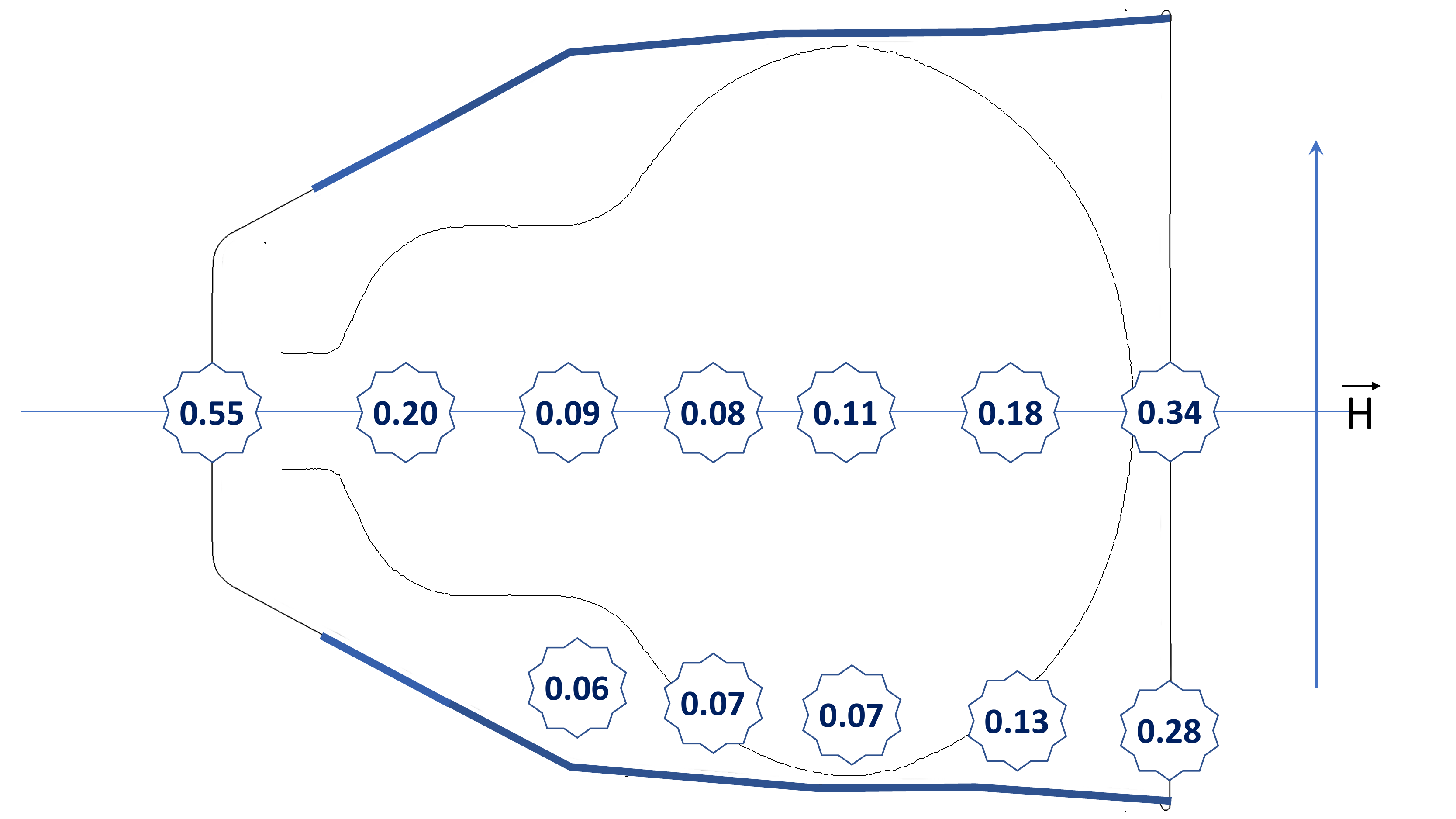}
\includegraphics[width=0.49\textwidth]{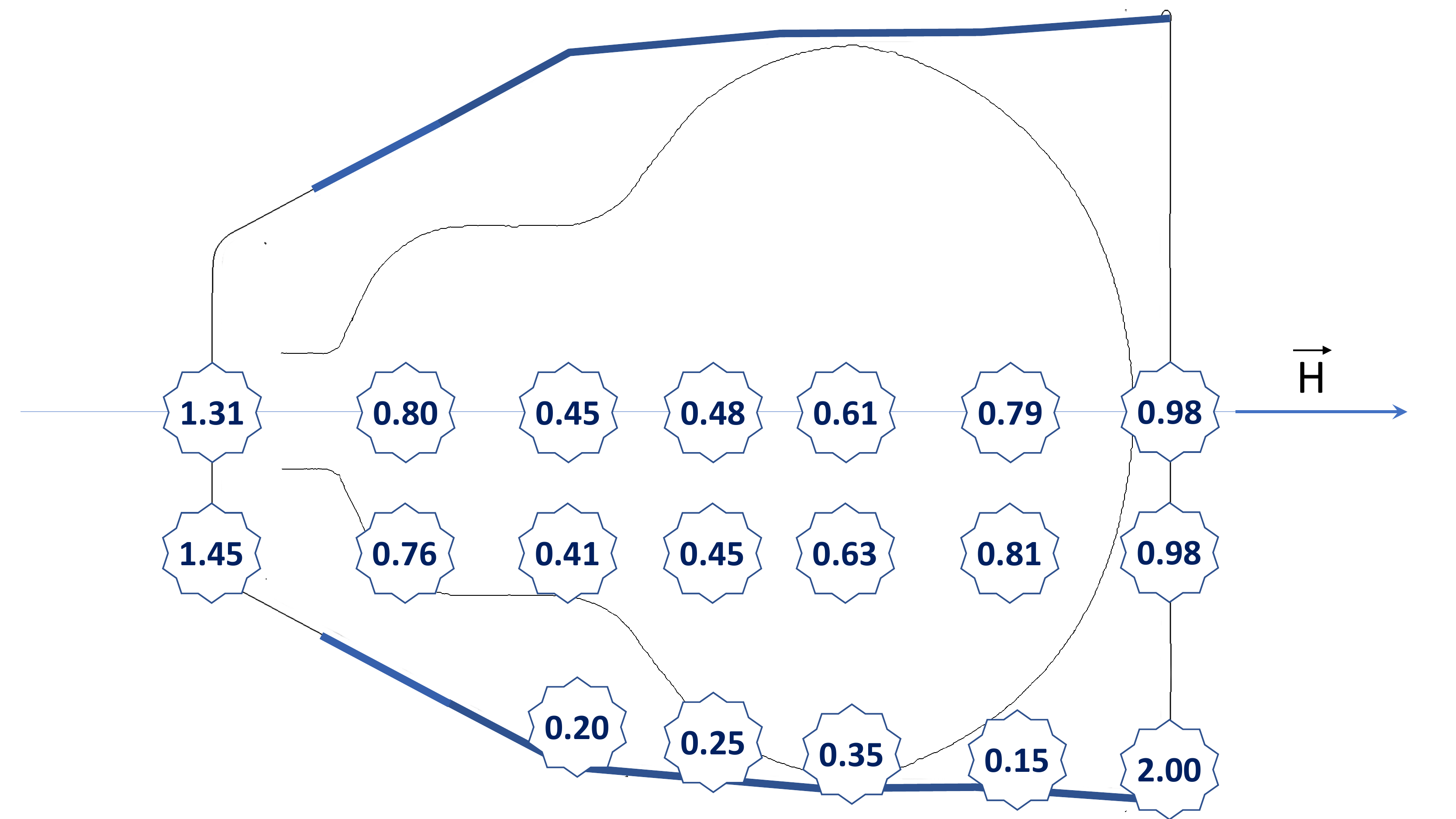}
\caption{Suppression of the weak magnetic field for two mutually orthogonal components: perpendicular to the PMT axis and along the PMT axis, or z-axis. The position of the PMT glass bulb is shown for reference, no PMT was physically present at the time of measurement. The location of the magnetic shield in the cone is shown with a thick line.}
\label{Suppression}
\end{figure}

The suppression of the weak magnetic field inside the shield is shown in Fig.~\ref{Suppression}. The unit of measurement is 50~$\mu$T (roughly corresponds to the EMF strength), the suppression coefficient is a fraction of the total field measured at the given point inside the shield. 

Most critical for the normal PMT operation is the direction along one of the axes orthogonal to the PMT symmetry axis (z-axis in the chosen coordinate system, see Fig.~\ref{Axes}), namely the direction along the dynodes (y-axis), see~\cite{BrxShield}. Transport of electrons inside the PMT bulb is most sensitive to the field in this direction. The estimates of shield efficiency were performed for the field orientation along the y-axis. The estimates obtained correspond to the maximum possible influence of the magnetic field in the direction orthogonal to the PMT symmetry axis.

As one can see in Fig.~\ref{Suppression}, the shield suppresses the external field in the direction orthogonal to the symmetry axis by a factor of 3 to 12. This means that a large part of the working space inside the PMT volume is well protected against this component of the EMF. For the vertically installed PMTs, the residual sensitive component will range from 3 to 12~$\mu$T depending on the position inside the PMT bulb. The integral effect should be checked anyway, because a noticeable reduction of PMT efficiency is observed for uniform fields of above 5~$\mu$T.

The suppression coefficient along the PMT axis is significantly weaker, its maximum value close to the position in the PMT bulb center is just 2. Nevertheless, due to the much less pronounced sensitivity of PMTs to this component, we do not need its strong suppression.

Earlier measurements with 20" dynode PMTs of the JUNO detector showed a weak dependence on time characteristics (below $\pm$0.2 ns) in the external magnetic field of up to 20 $\mu$T~\cite{JUNO20inch}. Due to the fact that the magnetic shield suppresses the external field to this value, we focused on the study of light detection efficiency losses in external magnetic fields, anyway, we checked the influence of the magnetic field on the transit time spread.

The estimate of magnetic shield efficiency is obtained by comparing light detection efficiency in the absence of the magnetic field and when the magnetic field is applied. As reference values, we used  the measurements in the compensated magnetic field with the magnetic shield installed. The measurements with and without a magnetic shield in the compensated field allowed estimating the improvement of light collection due to the light concentrator. It is $(19.0\pm0.7)$\% $\footnote{In the air. In experimental conditions, PMTs will be placed in a medium with a higher refraction coefficient (in water), this leads to improvement of light detection with a spherical photocathode. Correspondingly, the improvement of light collection efficiency could be lower in water.}$. 

Tests of the magnetic shield were performed at a special laboratory equipped with Helmholtz coils for EMF compensation~\cite{Anfimov}. Changing the orientation of a PMT in space and adjusting currents in compensation coils, it was possible to apply fields from 0 to 50 $\mu$T along the PMT axis or in the orthogonal direction. The light field was created by a short-pulse laser. The efficiency of light detection was estimated by an average number of photoelectrons (p.e.) registered for one trigger of the laser. In order to provide the stability of the system, a special care was taken. The stability was estimated by taking data in the same conditions at different periods of time. We used in total 15 datasets of 30000 events in each, the variance of central values was found to be $\delta \mu=0.026$ p.e. while the average p.e. count was $\overline{\mu}=1.63$ with expected statistical uncertainty of $\sigma_{\mu}=0.012$ p.e. Using these values, the systematics of the measurements can be estimated  $\mu=0.023$ p.e. or 1.4\%.

When the maximum field of 40 $\mu$T is applied along the most sensitive y-axis, the average count of photoelectrons drops by a factor of 2 ($\mu$=0.88 p.e., with a reference value in the compensated field of $\mu$=1.85 p.e.). The electronic gain of a multiplier estimated from charge histograms decreases by a factor of 1.4. The decrease in light detection efficiency was not so pronounced for the same field applied along the x-axis: light detection efficiency decreased by a factor of 1.09, and the electronic gain decreased by a factor of 1.14. Uncertainties of the values quoted are at a level of 2\% with dominating systematics.

Light detection efficiency of the PMT with the magnetic shield in the maximum field of 40~$\mu$T coincides with reference values within the error of measurements. 

The transit time spread (TTS) was checked for a field of up to +/-40~$\mu$T directed along the x-, y- and z-axes. In the worst case, the TTS for the PMT without a magnetic shield increases by about 30\% in the presence of the magnetic field (from 0.9 to 1.2~ns for the PMT under tests), the TTS for the PMT equipped with the magnetic shield is about 0.9~ns for all possible directions of the magnetic field, which is consistent with the value measured in the absence of the magnetic field within the precision of measurement. 

We can state that the magnetic shield developed at our laboratory provides practically ideal protection of PMTs in arbitrarily directed external magnetic fields of up to 40$\mu$T. 

\section{Conclusions}

There is no available space for installation of EMF compensation coils due to the configuration of the OSIRIS facility, and the detector PMTs will operate in the non-compensated EMF. We have developed open magnetic field shields using soft magnetic materials on the basis of amorphous alloys. The laboratory tests demonstrated very good efficiency of these magnetic shields for PMTs arbitrarily oriented in the EMF. Taking into account that the detector PMTs will not operate at extreme values of magnetic fields applied during the tests, the expected performance will approach an ideal one in a completely compensated field. 
Amorphous alloys reduce the amount of material needed for manufacturing due to their extremely high magnetic permeability. All the construction materials were tested for the content of radioactive long-lived natural isotopes U, Th and K. The total radioactivity of the magnetic shields produced on the basis of carbon fibre satisfies the low-background experiment requirements. The upper part of the magnetic shield is used as a light concentrator providing an additional 19\% in light collection efficiency in the air.

\section{Acknowledgements}
This work was supported by the Ministry of Science and Higher Education of the Russian Federation, Contract 075-15-2020-778 of the Program of Major Scientific Projects within the National Project Science.
We are grateful to N. Mazarskaya for her valuable help in preparing the manuscript.

\end{document}